\documentclass{JHEP3}
\usepackage{latexsym}

\def\AdSs5{$AdS_5$}
\def\AdS5s5{$AdS_5 \times S^5$}

\def\RR{{$R\otimes R$}}

\def\calP{{\cal P}}

\def\calN{{\cal N}}
\def\Tr{\mbox{Tr}}

\newcommand{\ie}{{\it i.e.~}}

\newcommand{\be}{\begin{equation}}
\newcommand{\ee}{\end{equation}}
\newcommand{\ba}{\begin{eqnarray}}
\newcommand{\ea}{\end{eqnarray}}

\newbox\SlashedBox
\def\fs#1{\setbox\SlashedBox=\hbox{#1}
\hbox to
0pt{\hbox to 1\wd\SlashedBox{\hfil/\hfil}\hss}{#1}}
\def\hboxtosizeof#1#2{\setbox\SlashedBox=\hbox{#1}
\hbox to
1\wd\SlashedBox{#2}}

\def\ms#1{\setbox\SlashedBox=\hbox{$#1$}
\hbox to 0pt{\hbox to
1\wd\SlashedBox{\hfil/\hfil}\hss}#1}

\def\t2{\tau_2}
\def\AdSS5{$AdS_5$}
\def\AdS5s5{$AdS_5
\times S^5$}

\def\calP{{\cal P}}

\def\IZ{\relax\ifmmode\mathchoice {\hbox{\cmss Z\kern-.4em
Z}}{\hbox{\cmss Z\kern-.4em Z}} {\lower.9pt\hbox{\cmsss Z\kern-.4em
Z}} {\lower1.2pt\hbox{\cmsss Z\kern-.4em Z}}\else{\cmss Z\kern-.4em
Z}\fi}

\def\Tr{{\rm Tr}}

\def\RR{{$R\otimes R$}}

\def\c1{{\chi^1}}

\def\N4{{\cal N}=4}
\def\half{{1\over 2}}
\def\nn{\nonumber}

\def\Tr{{\rm Tr}}
\def\x{{\bf x}}

\def\A{{\cal A}}

\def\ie{{\it i.e.}}
\def\half{{{1 \over 2}}}

\def\pmb#1{\setbox0=\hbox{#1}%
 \kern-.025em\copy0\kern-\wd0
 \kern.05em\copy0\kern-\wd0
 \kern-.025em\raise.0433em\box0 }
\font\cmss=cmss10
\font\cmsss=cmss10 at 7pt
\def\rlx{\relax\leavevmode}
\def\Cop{\relax\,\hbox{$\kern-.3em{\rm C}$}}
\def\Rop{\relax{\rm I\kern-.18em R}}
\def\Nop{\relax{\rm I\kern-.18em N}}
\def\Pop{\relax{\rm I\kern-.18em P}}

\def\Zop{\rlx\leavevmode\ifmmode\mathchoice{\hbox{\cmss Z\kern-.4em Z}}
 {\hbox{\cmss Z\kern-.4em Z}}{\lower.9pt\hbox{\cmsss Z\kern-.36em Z}}
 {\lower1.2pt\hbox{\cmsss Z\kern-.36em Z}}\else{\cmss Z\kern-.4em
 Z}\fi}

\def\m{{m}}

\def\xxx#1 {{hep-th/#1}}

\def\npb#1(#2)#3 { Nucl. Phys. {\bf B#1} (#2) #3 }
\def\rep#1(#2)#3 { Phys. Rept.{\bf #1} (#2) #3 }
\def\plb#1(#2)#3{Phys. Lett. {\bf #1B} (#2) #3}
\def\prl#1(#2)#3{Phys. Rev. Lett.{\bf #1} (#2) #3}
\def\physrev#1(#2)#3{Phys. Rev. {\bf D#1} (#2) #3}
\def\ap#1(#2)#3{Ann. Phys. {\bf #1} (#2) #3}
\def\rmp#1(#2)#3{Rev. Mod. Phys. {\bf #1} (#2) #3}
\def\cmp#1(#2)#3{Comm. Math. Phys. {\bf #1} (#2) #3}
\def\mpl#1(#2)#3{Mod. Phys. Lett. {\bf #1} (#2) #3}
\def\ijmp#1(#2)#3{Int. J. Mod. Phys. {\bf A#1} (#2) #3}

\title{D-brane interactions in type IIB plane-wave background}
\author{Oren Bergman \\ Department of Physics \\
Technion, Israel Institute of Technology, Haifa 32000, Israel\\
E-mail: \email{bergman@physics.technion.ac.il}}
\author{Matthias R. Gaberdiel \\ Department of Mathematics \\
King's College London, Strand, London WC2R 2LS, UK \\
E-mail:
\email{mrg@mth.kcl.ac.uk}}
\author{Michael B. Green \\
Department of Applied Mathematics and Theoretical Physics \\
Wilberforce Road, Cambridge CB3 0WA, UK \\
E-mail:
\email{M.B.Green@damtp.cam.ac.uk}}

\abstract{The cylinder diagrams that determine the static
interactions between pairs of D$p$-branes in the type IIB plane wave
background are evaluated.  The resulting expressions are elegant
generalizations of the flat-space formulae that depend on the value of
the Ramond-Ramond flux of the background in a non-trivial manner.
The closed-string and open-string descriptions consistently transform into
each other under a modular transformation only when each of the
interacting  D-branes separately preserves half the supersymmetries. These
 results are derived for configurations of euclidean signature
 D$(p+1)$-instantons but also generalize to lorentzian signature D$p$-branes.}

\preprint{ KCL-MTH-02-11 \\ DAMTP-2002-48 \\ {\tt hep-th/0205183}}

\begin{document}

\section{Introduction}
\label{intro}

Type IIB superstring theory in a plane-wave (or $pp$-wave) background
can be formulated as a free two-dimensional field theory, at least in
the light-cone gauge, which makes this an interesting model for
testing general string theory ideas.  This background, which has a
non-zero flux of the Ramond--Ramond (RR) five-form field strength
preserves the maximal number of 32 supersymmetries, which is the same
amount of supersymmetry as in the flat background as well as in
$AdS_5\times S^5$.  Indeed, the plane-wave background can be obtained
as a Penrose limit of $AdS_5\times S^5$ \cite{hulletal} and it has
been shown that the AdS/CFT correspondence has a corresponding limit
\cite{maldetal}.  In particular, the excited string states have been
shown to correspond to a class of gauge invariant operators in the
large $N$ limit of $\calN=4$ supersymmetric $SU(N)$ Yang--Mills
theory.  Such a correspondence is very difficult to demonstrate in the
original $AdS_5\times S^5$ background since the string theory has not,
so far, proved tractable in that case.  The bosonic isometry group of
the plane-wave background has 30 generators -- the same number as for
$AdS_5\times S^5$.  Although the metric has an $SO(8)$ isometry group
this is broken to $SO(4) \times SO(4)$ by virtue of the background
$F_5$ flux, which distinguishes the directions $x^I$ with $I= 1,2,3,4$
from the directions with $I= 5,6,7,8$.  In this notation the
light-cone directions are $x^\pm = (x^9 \pm x^0)/\sqrt{2}$, where
$x^0$ is time-like.

Although the spectrum of states is easy to determine in the light-cone
gauge \cite{met,mt}, this parameterisation has some awkward features.
In particular, in this background the light-cone gauge world-sheet
theory is not conformally invariant since the world-sheet fields are
massive\footnote{A covariant version of this string theory has been
recently proposed in \cite{berk}.}.  As a result various quantities,
notably the string interactions, are difficult to evaluate in the
$pp$-wave background.

Even though the interactions between string states are difficult to
determine, it is relatively straightforward to determine the static
interactions between pairs of D-branes in the type IIB plane-wave
background. This is the subject of this paper. It has been shown that
the only supersymmetric D$p$-branes are those with $p=3$, $5$ and $7$
oriented in a particular manner with respect to the eight directions
transverse to the light-cone directions \cite{dab}.  As in the flat
space case, it is also possible to consider D$(p+1)$-instantons where
the light-cone directions are transverse to the euclidean
$(p+1)$-dimensional world-volume \cite{gg}.   In that case the
supersymmetric values of $p$ are $p=1$, $p=3$ and $p=5$ \cite{bp},
again with appropriate  orientations with respect to the transverse
coordinates\footnote{For convenience we shall generically refer to
either of these classes of objects as `D-branes' in this paper.}.
As will be explained later, these D-branes are generally not flat in
this parameterisation.  The parameterisation in which D-branes are flat
is one in which the transverse components of the metric are not
constant in the light-cone time coordinate.  It is therefore
inappropriate for a hamiltonian analysis.

In this paper we will consider the cylindrical world-sheet diagrams
that describe the exchange of closed strings, paralleling the
discussion of the interaction energy between D-branes in flat space.
We will see that the requirement that the expression for the cylinder
diagram has a consistent interpretation in terms of a loop of open
string restricts the spectrum of D$p$-branes and D$(p+1)$-instantons
to those that have been shown to be supersymmetric.  The expressions
for the cylinder interactions are elegant generalisations of the
flat-space expressions, which depend explicitly on the background
Ramond--Ramond flux.  They transform covariantly under the particular
modular transformation that relates the closed-string and open-string
channels, \ie\ the $S$ transformation that interchanges the
world-sheet space and time coordinates.
\smallskip

The paper is organised as follows. In the following subsection we
review briefly the type IIB light-cone gauge
$pp$ wave background and fix our notation. The
boundary states for the D$(p+1)$-instantons are defined in
section~2. In section~3 we analyse the static interaction between two
such D-branes which is determined, in leading perturbative approximation,
by a cylinder world-sheet diagram.  In the case of the flat-space theory,
which is reviewed in section~3.1, this interaction is given by a
ratio of powers of certain standard functions, $f_i$  ($i=1,2,3,4$).
These functions transform in a particular manner under the $S$ modular
transformation so that the cylindrical world-sheet can be described
either in terms of the exchange of a closed string between the
D-branes or as a vacuum loop of open string with end-points on the two
D-branes.  In section~3.2 we will describe generalisations of
$f_i$ to the interaction between a pair of D$(p+1)$-instantons in
the $pp$-wave background.  These generalized functions, $f_i^{(m)}$,
depend on the value of the five-form field strength, $\mu$.
We also comment on the physical interpretation of these diagrams in
section~3.4. Section~4 deals with the description of D$p$-branes with
lorentzian signature world-volumes and
section~5 contains some conclusions.  In appendix~A the $S$
transformation of the $f^{(m)}_i$ functions are derived, while
appendix~B contains some technical details of the cylinder
calculation.

\subsection{Notation and review}
\label{notation}

The $pp$-wave background is a ten-dimensional space-time with metric,
\be
ds^2 = 2 dx^+ dx^- - \mu^2 x^I x^I dx^+ dx^+ + dx^I dx^I \, ,
\label{metpp}
\ee
where $x^{\pm} =  (x^9 \pm x^0)/\sqrt{2}$ and $I=1,\dots,8$.
The five-form \RR\ field strength is given by
\be
F_{+ 1234} = F_{+ 5678} = 2\, \mu\, ,
\label{fback}
\ee
where $\mu$ is a constant.
This background possesses a thirty-dimensional isometry group that
contains $SO(4) \times SO(4)$, together with ten translations and the
eight boosts generated by $J^{+I}$.  The supersymmetries extend this
to a supergroup.

In light-cone gauge string theory the
time-like parameter of the world-sheet is defined by the choice of
parameterisation,
\be
x^+ = 2 \pi\, \alpha' \, p^+ \tau\, ,
\label{lcdef}
\ee
where the constant $p^+$ is the $+$ component of the momentum density.
This commutes with all the
other isometries, in contrast to the flat case where there is an
isometry generated by
$J^{-I}$ which does not commute with $p^+$.

In the following we shall follow many of the conventions of
\cite{mt,bp} (although we shall use $SO(8)$ notation
and choose $\alpha'=1$).
The light-cone lagrangian in the plane-wave background describes
eight massive free scalar and eight massive free fermion fields,
\be
{\cal L} = {1\over 4\pi} \left( \partial_+ x^I \partial_- x^I
- m^2 (x^I)^2  \right)
+ {i \over 2\pi}\left(S^a\partial_+ S^a + \tilde S^a \partial_- \tilde S^a
- 2m\, S^a\, \Pi_{ab} \, \tilde S^b \right)\, ,
\label{lcact}
\ee
where $S^a$ and $\tilde S^a$ are $SO(8)$ spinors of the same chirality
and $\Pi = \gamma^1 \bar\gamma^2\gamma^3 \bar\gamma^4$. The mass parameter
$m$ is defined by $m=2 \pi p^+ \mu$. The $8\times 8$ matrices,
$\gamma^I_{a\dot b}$ and $\bar\gamma^I_{\dot ab}$, are the
off-diagonal  blocks of the $16\times 16$ $SO(8)$ $\gamma$-matrices
and couple $SO(8)$ spinors of opposite chirality.  The presence of
$\Pi$ in the fermionic sector of the lagrangian breaks the symmetry
from $SO(8)$ to $SO(4)\times SO(4)$.  As noted earlier, the mass terms
in the action (\ref{lcact}) are not conformally invariant.

The modes of the transverse bosonic
coordinates, $x^I$, of a string in type IIB $pp$-wave light-cone gauge
string theory  \cite{met} are $\alpha^I_k$ and $\tilde\alpha^I_k$, where
$I$ is a vector index of $SO(8)$, and $k\in\Zop$ with $k\ne 0$. These
modes satisfy the commutation relations
\be
[\alpha^I_k,\alpha^J_l]  = \omega_k \, \delta^{IJ}\, \delta_{k,-l}
\,, \qquad
[\alpha^I_k,\tilde\alpha^J_l]  = 0 \, , \qquad
[\tilde\alpha^I_k,\tilde\alpha^J_l] = \omega_k\, \delta^{IJ}\,
\delta_{k,-l} \,,
\label{boscom}
\ee
where
\be
\omega_k = {\rm sign}(k)\, \sqrt{k^2 + \m^2}\qquad |k|>0\, .
\label{omegadef}
\ee
In addition there are the bosonic zero modes that describe the
centre of mass position $x^I_0$ and some generalised momentum $p^I_0$
with
\be
[p_0^I,x_0^J] = - i \delta^{IJ} \,.
\label{boszero}
\ee
It is convenient to introduce the creation and annihilation operators
\be
a^I_0 = {1\over \sqrt{2\m}} \bigl(p_0^I + i \m x_0^I\bigr) \,, \qquad
\bar{a}^I_0 = {1\over \sqrt{2\m}} \bigl(p_0^I - i \m x_0^I\bigr) \,,
\label{creani}
\ee
in terms of which (\ref{boszero}) is then simply
\be
[\bar{a}_0^I,a_0^J] = \delta^{IJ}\,.
\label{creanii}
\ee
As explained in \cite{mt}, $p^I_0$ does not simply correspond to
the usual momentum along the $x^I$ direction, but rather is a linear
combination (with coefficients depending on $\mu$) of the momentum
components $p^I$ and $p^+$.  More explicitly, there are eight
isometries associated with the generators
\be
P^I =  \int_0^1 d\sigma \left(\cos (\mu x^+) \calP^I
+ 2\pi p^+ \mu \sin (\mu x^+) x^I \right) \equiv p_0^I \, ,
\label{transisom}
\ee
where ${\cal P}^I = \dot x^I \sim {\delta / \delta x^I}$ and
$p^+ \sim {\delta / \delta x^-}$. We see from this expression that
the isometries associated with $P^I$ combine displacements in the
$x^I$ directions with displacements in the $x^-$ direction.

The fermionic modes  $S^a_k$ and $\tilde{S}^a_k$, where $a$ is a
spinor index of $SO(8)$, and $k\in\Zop$, satisfy the anti-commutation
relations
\be
\{ S^a_k,S^b_l\} = \delta^{ab}\delta_{k,-l} \,,  \qquad
\{ S^a_k,\tilde{S}^b_l\}  = 0 \,, \qquad
\{\tilde{S}^a_k,\tilde{S}^b_l\}  = \delta^{ab}\delta_{k,-l} \,.
\label{fermcom}
\ee
It is convenient to introduce the zero-mode combinations,
\be
\theta^a_0 = {1\over \sqrt{2}}(S^a_0 + i \tilde{S}^a_0)
\,, \qquad \bar\theta^a_0 = {1\over \sqrt{2}}(S^a_0 - i \tilde{S}^a_0)
\,,
\label{thetadef}
\ee
and further
\ba
\theta_R & =& \half (1 + \Pi) \theta_0 \,,  \qquad
\bar\theta_R = \half (1 + \Pi) \bar\theta_0 \,, \nn\\
\theta_L & =& \half (1 - \Pi) \theta_0 \,, \qquad
\bar\theta_L = \half (1 - \Pi) \bar\theta_0 \,.
\label{thetarl}
\ea
The  light-cone hamiltonian $H$ for the string in the plane-wave
background is given by
\ba
2 \, p^+ H &=& \m \left(a^I_0\, \bar{a}^I_0
+ i\, S^a_0 \,\Pi_{ab}\, \tilde{S}^b_0 +4\right)
+ \sum_{k=1}^{\infty} \left[ \alpha^I_{-k}\alpha^I_k +
\tilde\alpha^I_{-k} \tilde\alpha^I_k
+ \omega_k \left(S^a_{-k} S^a_{k} + \tilde{S}^a_{-k} \tilde{S}^a_{k}
\right) \right]\nn\\
 &=& \m \left(a^I_0\, \bar{a}^I_0 + \theta_L^a \,\bar\theta_L^a +
\bar\theta_R^a \,\theta_R^a \right)
+  \sum_{k=1}^{\infty} \left[
\alpha^I_{-k}\alpha^I_k +
\tilde\alpha^I_{-k} \tilde\alpha^I_k
+ \omega_k \left(S^a_{-k} S^a_{k} + \tilde{S}^a_{-k} \tilde{S}^a_{k}
\right) \right]
\,.\nn\\
\label{lcham}
\ea
In the case $m=0$ this reduces to the usual light-cone gauge
hamiltonian in a flat background \cite{gs1}.   The normal ordering
has been chosen in (\ref{lcham}) with the understanding that
$\theta_L^a$ and $\bar\theta_R^a$ are creation operators while
$\bar\theta_L^a$ and $\theta_R^a$ are annihilation operators.

As is familiar from flat space, the space of
states is described by a tensor product of the space
generated by the bosonic modes and that generated by the fermionic
modes. The ground
state of the bosonic space, $|0\rangle_b$, is annihilated by the
modes $\bar{a}^I_0$ as well as $\alpha^I_k$ and $\tilde{\alpha}^I_k$
with $k>0$ and is non-degenerate since each of the
`zero modes' $a^I_0$ raises the energy by $\m$.  Likewise,
the non-degenerate ground state in the space spanned by the fermionic
operators, $|0\rangle_f$, is the state
annihilated by $\bar\theta_L^a$ and $\theta_R^a$, while
the creation operators $\theta_L^a$ and $\bar\theta_R^a$ raise the
energy by $m$.
When $p^+ \ne 0$ the total space (the tensor product of the bosonic
and fermionic spaces) contains a single unpaired bosonic ground state
of zero mass, while all excited states come in degenerate
boson-fermion pairs.  This is in accord with the expectations based on
supersymmetry. Recall that in the light-cone gauge the thirty-two
components of the supersymmetries decompose into `dynamical' and
`kinematical' components.
The `dynamical' supercharges, $Q^{\dot a}$  and
$\tilde Q^{\dot a}$ ($\dot a =1,\dots,8$), commute
with $H$.  These charges are therefore associated with sixteen
$x^+$-independent components
of a Killing spinor which can be identified with the
$\gamma_+ \gamma_- \epsilon$ projection of the 32-component
Killing spinor in equation (6.26) of \cite{ST}.  The `kinematical'
supercharges, $Q^a$ and $\tilde Q^a$ ($a =1,\dots,8$), do not commute
with $H$.  They are associated with the components
$\gamma_- \gamma_+ \epsilon$ in \cite{ST}, which are functions of
$x^+$.  It follows from the anti-commutation relation
$\{Q^{\dot a}, Q^{\dot b}\} \sim 2\delta^{\dot a \dot b}\, H +
{\cal O}(\mu)$
that the massive states ($H>0$) consist of pairs of fermions and
bosons related by the dynamical supersymmetries while the massless
state with $H=0$ is unpaired (the term of order $\mu$ is proportional
to the angular momentum generators $J_{IJ}$ and does not affect the
argument).   When $p^+=0$ the mass parameter
vanishes ($m=0$) and the massless spectrum degenerates to the
supersymmetric spectrum of the flat-space theory which has eight
massless fermions and eight massless bosons.

\section{Boundary states}
\label{boundstat}

To begin with we will pursue the boundary state
description of D-branes in terms of a light-cone gauge
defined with respect to a time-like direction that is transverse to
the $(p+1)$-dimensional world-volume of a D$(p+1)$-instanton.   We are
therefore assuming that the light-cone  directions,
$x^+ =   (x^9+x^0)/ \sqrt{2}$ and
$x^- =   (x^9-x^0)/ \sqrt{2}$ are transverse to the brane.
This parameterisation was used in the context of the bosonic
D-instanton in \cite{gone} and the type IIB D-instanton  in
\cite{gtwo} and was generalized to type II  D($p+1)$-instantons in
flat space-time in \cite{gg}.  It
provides a particularly efficient and manifestly supersymmetric
way of calculating the closed string propagating
in the cylinder diagram that describes the interactions between
D-branes. In this description
there are at most eight Neumann directions, so it is not possible to
describe the D$8$-brane (of the type IIA theory) or the D$9$-brane (of
the type IIB theory) in this light-cone formalism.  The boundary
states of D($p+1$)-instantons in the IIB plane-wave
background considered in \cite{bp},
$|\!| D(p+1),   p^+\,\rangle\!\rangle$, are closed-string states that
satisfy appropriate `gluing conditions' for a world-volume located at
$x^+=0$, with momentum $p^+$ (which is the Fourier conjugate of the
transverse coordinate $x^-$) and with  transverse position $\x=0$
(where $\x$ indicates the $(7-p)$ components of $x^I$ that satisfy
Dirichlet conditions).  The  latter condition is required by the
consistency conditions reviewed below.

The gluing conditions in the bosonic sector are taken to have the form
\ba
\left( \alpha^I_k - M_{IJ}\, \tilde\alpha^J_{-k} \right)
|\!| D(p+1), p^+\,\rangle\!\rangle & =& 0 \qquad k\in\Zop\,, \nn\\
\left( \bar{a}^I_0 - M_{IJ}\, a^J_0 \right)
|\!| D(p+1), p^+\,\rangle\!\rangle & = & 0 \,,
\label{boundarycond}
\ea
where $M_{IJ}$ is an $SO(8)$-matrix. These conditions are a simple
generalization of the conditions that apply in the flat case. We
shall restrict our attention to the case when $M_{IJ}$ is a diagonal
matrix with $\pm 1$ on the diagonal, $M_{IJ}=\epsilon_I\delta_{IJ}$.
The directions with $\epsilon_I=+1$ satisfy a Dirichlet boundary
condition with $x^I=0$, while those with $\epsilon=-1$ obey the
Neumann-like boundary condition $p^I_0=0$. By the construction of the
light-cone gauge, $x^+$ satisfies a Dirichlet boundary condition. The
boundary condition for $x^-$ is determined by the 
Virasoro constraints 
\ba\label{constraint}
{x'}^- & = & - {\dot{x}^I {x^\prime}^I \over 2\pi p^+} \cr
\dot{x}^- & = & - {\dot{x}^I \dot{x}^I +{x^\prime}^I {x^\prime}^I
                    - m^2 x^I x^I \over 4\pi  p^+}\,,
\ea
where 
the contributions from the fermions have been ignored, and 
the dot denotes the derivative with respect to $\tau$, while the
prime denotes the derivative with respect to the world-sheet spatial
coordinate $\sigma$. If $x^I$ satisfies a Dirichlet boundary condition
(at $\tau=\tau_0$), then ${x^\prime}^I=0$ at $\tau=\tau_0$, but if
$x^I$ satisfies $p^I_0=0$, then it follows from (\ref{transisom}) that
$\dot{x}^I = - m \tan(\mu x^+) x^I$ at
$\tau=\tau_0$. Thus, it follows from (\ref{constraint})
that $x^-$ satisfies a generalized Dirichlet condition that
relates it to $x^I$,
\be\label{diffeq}
{x'}^- = \sum_{I\in{\cal N}} {x^\prime}^I x^I \mu \tan(\mu x^+) \,,
\ee
where the sum is over all `Neumann' directions, \ie\ the
world-volume values of $I$ which have $p_0^I=0$. In light-cone gauge
$x^+$ is independent of $\sigma$, and (\ref{diffeq}) can be integrated
to give
\be\label{xminus}
x^- = \sum_{I\in{\cal N}} \half x^I x^I \mu \tan(\mu x^+) + c(\tau) \,,
\ee
at $\tau = \tau_0$.
The function $c(\tau)$  can be determined by considering the
$\tau$ derivative of this equation and comparing it with the second
equation of (\ref{constraint}), using the boundary condition on
$\dot{x}^I$, which gives
\be
\left.  \dot{c}(\tau)\right|_{\tau = \tau_0} =
   - \left.{1\over 4\pi p^+} \sum_{I\in {\cal N}} {x^\prime}^I
           {x^\prime}^I\right|_{\tau = \tau_0}
\,.
\label{dotxm}
\ee
For this to be consistent it is obviously important that the
right-hand side is independent of $\sigma$, so that it must be
true that
\be
\left. \sum_{I\in {\cal N}} {x^{\prime\prime}}^I
{x^\prime}^I\right|_{\tau = \tau_0} = 0 \,.
\label{sigind}
\ee
This follows upon using the equation of motion,
\be
{x^{\prime\prime}}^I = \ddot{x}^I + m^2 x^I
\label{eqmot}
\ee
and the boundary condition $\dot{x}^I = - m \tan(\mu x^+) x^I$.

It follows from (\ref{xminus})
that $x^-$ is a quadratic function of $x^I$ at the boundary
$\tau=\tau_0$.  The relevant D-branes are therefore not
flat in this coordinate frame, since the position of the D-brane
in the $x^-$ direction varies with the position on the world-volume of
the brane (that is described by $x^I$).  Furthermore, the curvature
depends on the value of  $x^+$, becoming infinite at the points
$x^+= (2n- 1)\pi /2\mu$,
where $n$ is an integer.  The same system may be considered in Rosen
coordinates \cite{met}\footnote{These are coordinates in which the
metric is given by
$ds^2 = 2 dx^+ dx^- + \cos(\mu x^+)\; dx^I dx^I$.}, in
which case these values of $x^+$ coincide with the points at
which the metric is singular.  The operator
$P^+\sim \delta/\delta x^-$ is an isometry that generates constant
shifts of $x^-$.

The boundary states are assumed, as in flat space \cite{gg}, to be
annihilated by half the `kinematical' light-cone supercharges,
\be\label{kinematical}
\left( Q_a + i \eta\, M_{ab}\, {\tilde Q}_b \right)
|\!| D(p+1),p^+,\eta\,\rangle\!\rangle = 0\,,
\ee
where the value of $\eta =\pm 1$ distinguishes a brane from an
anti-brane.
The simplest way to satisfy this condition is to assume that 
the gluing conditions for the fermions are given as 
\be
\left( S^a_k + i \eta \, M_{ab} \, \tilde{S}^b_{-k} \right)
|\!| D(p+1),p^+,\eta\,\rangle\!\rangle = 0\,, \qquad k\in\Zop \,.
\label{boundarycondii}
\ee
In the following we shall only consider boundary states that satisfy
this gluing condition. We shall furthermore be interested in boundary
states that preserves half the `dynamical' light-cone supercharges, 
\be
\left( Q_{\dot a} + i \eta\, M_{\dot a \dot b}\,
{\tilde Q}_{\dot b}\ \right)
|\!| D(p+1),p^+,\eta\,\rangle\!\rangle = 0\,.
\label{dynamical}
\ee
For example in the flat space case where
$M_{IJ}= \epsilon_I \delta_{IJ}$ with $\epsilon_{I}=\pm 1$, we can
choose $M_{ab}$ to be given by
\be
M_{ab} = \left( \prod_{\epsilon_I=-1} \gamma^I \right)_{ab}\,,
\label{simplecase}
\ee
where we have assumed that the number of directions with
$\epsilon_I=-1$ is even so that the product involves an even number of
$\gamma$ matrices. Then (\ref{dynamical}) is satisfied with
\be
M_{\dot{a}\dot{b}} =
\left( \prod_{\epsilon_I=-1} \gamma^I \right)_{\dot{a}\dot{b}}\,,
\label{Mdot}
\ee
where we have chosen the same ordering as in (\ref{simplecase}).

In the $pp$ wave background $M_{ab}$ can still be chosen as in
(\ref{simplecase}). However, now (\ref{dynamical}) will only be
satisfied (with $M_{\dot{a}\dot{b}}$ given as in (\ref{Mdot})) provided
that the number of Dirichlet directions among the last
four transverse coordinates ($x^5,\ldots,x^8$) differs from the
number of Dirichlet directions among the first four coordinates
$(x^1,\ldots,x^4$) by $\pm 2$ \cite{bp}.
 In particular, D$(p+1)$-instantons with $p=-1$
and $p=7$  cannot satisfy this criterion.  These results hold
more generally, when the matrices $M_{ab}$ and
$M_{\dot{a}\dot{b}}$ are allowed to be arbitrary.

A complementary analysis of the open-string spectrum of D$p$-branes
(\ie\ branes with lorentzian world-volumes) in \cite{dab} uses a
light-cone gauge in which $x^\pm$ are world-volume directions.  In
this case one finds that the open-string spectrum is only
supersymmetric provided that $p=3$, $5$ and $7$ (and the number of
Dirichlet directions among the last four transverse coordinates
differs again from the number of Dirichlet directions among the first
four coordinates  by $\pm 2$). These values of $p$ exceed those of the
euclidean instantonic branes by $2$ since the coordinates $x^\pm$ now
form part of the world-volume rather than the transverse space. We
shall come back to D$p$-branes with lorentzian world-volumes in
section~4.

Imposing the conditions (\ref{boundarycondii}) together with
(\ref{boundarycond}) determines  the boundary state up to an overall
multiplicative constant normalization,
\be
|\!| D(p+1), p^+, \eta\,\rangle\!\rangle = {\cal N}
\exp \left( \sum_{k=1}^{\infty}
{1\over \omega_k} M_{IJ} \alpha^I_{-k} \tilde\alpha^J_{-k}
- i \eta M_{ab} S^a_{-k} \tilde{S}^b_{-k} \right) \,
|\!| D(p+1), p^+,\eta\,\rangle\!\rangle_0\,,
\label{boundone}
\ee
where the ground state component is
\be
|\!| D(p+1),p^+,\eta\,\rangle\!\rangle_0 =
\left( M_{IJ} |I\rangle |J\rangle
+ i \eta M_{\dot{a} \dot{b}} |\dot{a}\rangle |\dot{b}\rangle \right)
\, e^{\half M_{IJ} a_0^I a_0^J} |0\rangle_b \,.
\label{boundzero}
\ee
Here the first bracket describes the `fermionic' part of the ground
state, while the second part describes the `bosonic' part of the
ground state. Using the relations
$\sqrt{2}\; S^a_0 |I\rangle =  \gamma^I_{a\dot{a}}|\dot{a}\rangle$
and
$\sqrt{2} \; S^a_0 |\dot{a}\rangle =  \gamma^I_{a\dot{a}}|I\rangle$
it is not difficult to check that (\ref{boundzero})  satisfies the
zero mode part of (\ref{boundarycondii}).
\smallskip

The interaction energy between a pair of D-branes can be expressed, to
lowest order in the string coupling, by the exchange of a closed
string between two boundary states.  In the case of instantonic branes
this diagram describes the action rather than the energy. In the
following we want to show that the above branes are actually
singled out in an alternative manner in that they are the only ones
that give consistent expressions for these cylinder diagrams.
In particular, we will demonstrate that they are the only
boundary states that lead to a closed-string tree amplitude that
gives rise, upon a modular transformation, to a consistent open-string
one-loop amplitude. In terms of the conformal field theory, this
condition is often referred to as the `Cardy condition'.

\section{Interactions between pairs of D-brane instantons}

In flat space the expression for the cylinder diagram  in light-cone
coordinates can be derived starting from a covariant expression in the
following manner. The cylinder diagram is the tree-level amplitude,
\be\label{tree}
\A_{Dp_1;Dp_2} (X^+,X^-,\x_1,\x_2) =
\langle\!\langle Dp_1,x_1^+,x_1^-,\x_1 |\!| \Delta
|\!| Dp_2,x_2^+,x_2^-,\x_2 \rangle\!\rangle\,,
\ee
where $\Delta$ is the covariant closed-string propagator and
$\x_1$ and $\x_2$ are the transverse positions of the branes
that are also separated in the light-cone directions by
$X^\pm = (x^\pm_2 -x^\pm_1)$.
It is convenient to evaluate this in terms of boundary states that are
eigenstates of light-cone momentum, in which case the above expression
can be written as
\ba\label{fourier}
&& \hspace*{-0.5cm} \A_{Dp_1;Dp_2} (X^+,X^-,\x_1,\x_2) \nn\\
&= & \int {dp^+ dp^-\over 2\pi i} \, e^{i p^+ X^- + i p^- X^+}
\, \langle\!\langle Dp_1,-p^-,-p^+,\x_1 |\!|
\left({1\over   p^+ (p^- + H) } \right)
|\!| Dp_2,p^-,p^+,\x_2 \rangle\!\rangle \nonumber \\
& = &  \int_{-\infty}^\infty dp^+ e^{i p^+ X^-}\,
{\theta(p^+) \over  p^+} \,
\langle\!\langle Dp_1,-p^+,\x_1 |\!|
e^{-i H X^+} |\!| Dp_2,p^+,\x_2 \rangle\!\rangle\, ,
\ea
where $H$ is the closed-string light-cone hamiltonian,
and $\theta(p)$ is the step-function that is zero for negative $p$,
and one for positive $p$. We have used the usual causal $i\epsilon$
prescription in which the denominator is actually
$p^+(p^- + H) - i\epsilon = p^+(p^- + H - {i\epsilon/ p^+})$.
Since $X^+>0$, the only poles in $p^-$ that contribute have
Im$(p^-)>0$. This requires $p^+>0$, thus giving rise to the
$\theta(p^+)$ term.

Equation (\ref{fourier}) has the form
\be\label{covariant}
\A_{Dp_1;Dp_2}(X^+,X^-, {\x_1},\x_2) =
\int_0^\infty {dt\over t}\, e^{- {X^+ X^- \over 2 \pi t}} \,
\tilde \A_{Dp_1;Dp_2} (t,\x_1,\x_2)\,,
\ee
where $\tilde \A$ is the overlap
\be\label{Atilde}
\tilde \A_{Dp_1;Dp_2} (t,\x_1,\x_2)=
 \langle\!\langle Dp_1,-p^+,\x_1  |\!| e^{-2 \pi t H p^+}
|\!| Dp_2, p^+, \x_2 \rangle\!\rangle\,,
\ee
and $t$ is defined as
\be\label{tdef}
t = i {X^+\over 2 \pi p^+}\,.
\ee
In writing (\ref{covariant}) a customary Wick rotation, replacing
$\tau$ by $t=i\tau$, has been performed. The resulting world-sheet
theory is then euclidean.

In the usual flat space case   $\tilde{\A}$ can be
expressed in terms of the $f$-functions,
\ba
f_1(q) & =& q^{{1\over 24}} \prod_{n=1}^{\infty} (1-q^n) \,, \nn\\
f_2(q) & = &\sqrt{2}\,
 q^{{1\over 24}} \prod_{n=1}^{\infty} (1+q^n) \,,\nn\\
f_3(q) & =& q^{-{1\over 48}} \prod_{n=1}^{\infty}
\Bigl(1+q^{(n-1/2)}\Bigr) \,,\nn\\
f_4(q) & =& q^{-{1\over 48}} \prod_{n=1}^{\infty}
\Bigl(1-q^{(n-1/2)}\Bigr) \,.
\label{ffunc}
\ea
Here, as in the following, we have set $q=e^{-2\pi t}$. For example,
the amplitude of two identical BPS D-branes vanishes, and the
configuration of a BPS D$(p+1)$-instanton with an anti-BPS
D$(p+1)$-instanton gives rise to\footnote{We adopt a convention here
and in the following that the labels $Dp$ and $\bar Dp$ on $\A$ and
$\tilde \A$ are the same as if we were describing D$p$-branes with
lorentzian world-volumes.}
\be
\tilde{\A}_{Dp;\bar Dp}(t, \x_1,\x_2) =
 \left({f_2(q) \over f_1(q)}\right)^8\,
t^{{p-7\over 2}}\, e^{-{(\x_1 - \x_2)^2\over 2\pi t}}\,.
\label{bpsanti}
\ee
The power of $t^{{(p-7)/2}}$ in (\ref{bpsanti}) can be associated
with a factor that comes from the Fourier transform from momentum
space to position space transverse to the world-volume of the brane,
$\int d^{7-p}{\bf k} \; e^{i {\bf k}\cdot (\x_2-\x_1)}\,
e^{-\half \pi t {\bf k}^2}$.

One of the important consistency checks of D-branes is the
condition that requires that the above closed-string boundary
state overlap
can be re-interpreted as a one-loop open-string vacuum amplitude.
In order to relate the closed-string and the open-string points of
view, one has to exchange the (euclidean) world-sheet space and
time coordinates. This amounts to a conformal transformation,
replacing $t$ by $\tilde{t}={1/ t}$,
or $q$ by $\tilde{q}=e^{-2\pi/t}$. The resulting open string is then
formulated in a non-standard light-cone gauge where
$2 \pi p^+=X^+ \tilde t$.  The open-string hamiltonian in this
gauge is defined by,
\be
{X^+ \over 2\pi}\; H^{open} =
{1\over 2\pi} \sum_{I=p+2}^8 (x_1^I - x_2^I)^2
+  \pi \sum_{I=1}^{p+1} (k^I)^2 + 2 \pi N^{open}\,.
\label{hdef}
\ee
Here $N^{open}$ is the total level number of the bosonic and fermionic
open-string oscillators, including the usual normal ordering
constants, and $k^I$ are the momenta in the Neumann
directions $I= 1,\ldots, p+1$.

In the flat space case the consistency between the closed and open
string sectors is satisfied because the $f$-functions satisfy the
transformation properties
\be
f_1(q) = t^{-{1\over 2}} f_1(\tilde{q})  \,, \qquad
f_2(q) = f_4(\tilde{q}) \,, \qquad f_3(q) = f_3(\tilde{q}) \,.
\label{fmode}
\ee
For example, for the configuration of the BPS D$(p+1)$-instanton with
an anti-BPS D$(p+1)$-instanton discussed before, the function
$\tilde{\A}$ becomes
\be
\tilde{\A}_{Dp;\bar Dp}(t, \x_1,\x_2)
=  \left({f_4(\tilde{q}) \over f_1(\tilde{q})}\right)^8\,
\tilde{t}^{-{p+1\over 2}}\, e^{-{(\x_2-\x_1)^2\over 2\pi}\tilde{t}}
\, .
\label{bpsantitrans}
\ee
Using (\ref{covariant}) the  position-space cylinder amplitude
 is then given by
\be
\A_{Dp;\bar Dp} (x^+,x^-,\x_1,\x_2) =  \int_{0}^{\infty}
{d \tilde{t}\over \tilde{t}} \;
e^{-{X^+ X^- + (\x_1 - \x_2)^2 \over 2\pi} \tilde{t}} \;
\tilde{t}^{-{p+1\over 2}} \,
\left({f_4(\tilde{q}) \over f_1(\tilde{q})}\right)^8\,.
\label{oneloo}
\ee
This is to be compared with the open-string one-loop diagram, which
can be written as the integral over the circumference of the
cylinder of  a trace over open-string states,
\ba
Z_{Dp;\bar Dp}(X^+,X^-,\x_1,\x_2) &=&\int_0^\infty
{dp^+ \over p^+} \,e^{-p^+ X^- }\,
\tilde{Z}_{Dp;\bar Dp}(\tilde{t},\x_1,\x_2) \nn\\
&=& \int_0^\infty
{d\tilde{t} \over \tilde{t}} \,e^{-{X^+ X^- \over 2 \pi} \tilde{t}}\,
\tilde{Z}_{Dp;\bar Dp}(\tilde{t},\x_1,\x_2) \,,
\label{oneloop}
\ea
where
\be
\tilde{Z}_{Dp;\bar Dp}(\tilde{t},\x_1,\x_2) =  \Tr\, e^{-H^{open}\; p^+}=
\Tr\, e^{-{X^+ \over 2\pi}H^{open}\; \tilde t}\,,
\label{openstring}
\ee
and $H^{open}$ is defined in (\ref{hdef}).
 The zero-mode part of this trace is equivalent to integration over the
momenta in the Neumann directions and gives rise to a factor of
$\tilde{t}^{-{(p+1)/2}}$ in (\ref{oneloo}).
The open-string oscillators in the trace give a
factor $(f_4(\tilde{q})/f_1(\tilde{q}))^8$, while the
first term in $H^{open}$ gives rise to a factor of
$\exp\left({-(\x_1 - \x_2)^2 \tilde t / 2\pi}\right)$.
The expression (\ref{oneloop}) is therefore the same as (\ref{oneloo}).
From the point of view of the open-string
description, this diagram describes the free energy of open strings
whose endpoints lie on the two D-branes.

Provided that the brane and the anti-brane are sufficiently
separated, the integral defining the cylinder amplitude
(\ref{covariant}) converges. On the other hand, if the separation of
the branes becomes sufficiently small the diagram diverges, reflecting
the tachyonic instability of the system. In either case, the
consistency of the boundary state description requires that the two
expressions, (\ref{Atilde}) and (\ref{bpsantitrans}), agree, which
reduces to the `conformal field theory condition' that relates the
integrands,
\be\label{cftcond}
\tilde{\A}_{Dp_1;Dp_2}(t,\x_1,\x_2)
= \tilde{Z}_{Dp_1;Dp_2}(\tilde{t},\x_1,\x_2) \,.
\ee
This is the condition we shall analyse in the case of the $pp$-wave
background
in the following.

\subsection{The modified $f$-functions and their modular properties}

As we shall explain in more detail below, for the case of the
$pp$-wave background the cylinder amplitudes $\tilde{\A}$ or
$\tilde{Z}$ involve non-trivial deformations  of the $f$-functions
(\ref{ffunc}) which are given by the following expressions,
\ba
f_1^{(\m)}(q) & =& q^{-\Delta_\m} (1-q^\m)^{{1\over 2}}
\prod_{n=1}^{\infty} \left(1 - q^{\sqrt{\m^2+n^2}}\right) \,,
\label{fdef}\\
f_2^{(\m)}(q) & = &q^{- \Delta_\m} (1+q^\m)^{{1\over 2}}
\prod_{n=1}^{\infty} \left(1 + q^{\sqrt{\m^2+n^2}}\right) \,,
\label{f2def}\\
f_3^{(\m)}(q) & =& q^{-\Delta^\prime_\m}
\prod_{n=1}^{\infty} \left(1 + q^{\sqrt{\m^2+(n-1/2)^2}}\right) \,,
\label{f3def}\\
f_4^{(\m)}(q) & =& q^{-\Delta^\prime_\m}
\prod_{n=1}^{\infty} \left(1 - q^{\sqrt{\m^2+(n-1/2)^2}}\right) \,,
\label{f4def}
\ea
where $\Delta_\m$ and $\Delta^\prime_\m$ are defined by
\ba
\Delta_\m & =& -{1\over (2\pi)^2} \sum_{p=1}^{\infty}
\int_0^\infty ds \, e^{-p^2 s} e^{-\pi^2 \m^2 / s} \,, \cr
\Delta^\prime_\m & =& -{1\over (2\pi)^2} \sum_{p=1}^{\infty}
(-1)^p \int_0^\infty ds \, e^{-p^2 s} e^{-\pi^2 \m^2 / s}\,.
\label{Deltadef}
\ea
The quantities $\Delta_m$ and $\Delta^\prime_m$ are   the Casimir
energies of a single (two-dimensional) boson of mass $\m$ on a
cylindrical world-sheet with periodic and anti-periodic boundary
conditions, respectively. For $\m=0$, $\Delta_\m$ and
$\Delta^\prime_\m$ simplify to the usual flat-space values,
\ba
\Delta_0 & =& -{1\over (2\pi)^2} \sum_{p=1}^{\infty} {1\over p^{2}} =
- {1\over 24} \,, \cr
\Delta^\prime_0 & =& -{1\over (2\pi)^2}
\sum_{p=1}^{\infty} {(-1)^p \over p^{2}} =  {1\over 48}\,.
\label{mzero}
\ea
Thus $f_2^{(\m)}(q)$, $f_3^{(\m)}(q)$ and $f_4^{(\m)}(q)$ simply
reduce to the standard $f_2(q)$, $f_3(q)$ and $f_4(q)$ functions as
$\m\rightarrow 0$.

Using the same definitions as before for $q$ and $\tilde{q}$, these
functions transform as follows under the $S$ modular transformation
\be
f_1^{(\m)}(q)  = f_1^{(\widehat{\m})}(\tilde{q}) \,, \qquad
f_2^{(\m)}(q)  = f_4^{(\widehat{\m})}(\tilde{q}) \,,\qquad
f_3^{(\m)}(q)  = f_3^{(\widehat{\m})}(\tilde{q}) \, ,
\label{beautiful}
\ee
where
\be
\widehat{\m}=\m\, t\,.
\label{mhat}
\ee
These striking identities will be proven in appendix~A. The
definition of $\m$ depends on the choice of light-cone gauge
since $\m=2\pi \mu p^+$. In the Wick rotated theory $t={X^+/ 2 \pi p^+}$,
and thus
\be\label{mhatinter}
\m = 2\pi \mu p^+ \,, \qquad \widehat{\m} = \mu X^{+} \,.
\ee
Therefore $\widehat{\m}$ is the analogue of $m$ in the non-standard
open-string light-cone gauge (where the r\^oles of $X^+$ and $2 \pi p^+$
are reversed). For future reference we also note that
\be
e^{-2\pi m} = e^{-2\pi \tilde t \widehat m} =
\tilde q^{\,\widehat{\m}}\,  .
\label{tildhat}
\ee

In the limit $\m\rightarrow 0$ the second and third equations in
(\ref{beautiful}) reproduce the second and third identities of
(\ref{fmode}). The identity for   $f_1$ (or
$\eta$) can also be derived from the first equation of
(\ref{beautiful}).  This follows since both sides of
the first equation tend to zero as $\m\rightarrow 0$ since
$(1-q^\m)=2\pi t \m + {\cal O}(\m^2)$ and
$(1-\tilde{q}^{\,\widehat{\m}})=2\pi \m + {\cal O}(\m^2)$. Thus,
after dividing the first equation by $\m$ the limit
$\m\rightarrow 0$ becomes,
\be
f_1(q) = t^{-\half} f_1(\tilde{q}) \,,
\label{etatrans}
\ee
reproducing the first equation of (\ref{fmode}).

{}From a conformal field theory point of view, the transformation rule
$m \to \widehat{m}=mt$ also has a simple interpretation. The
characters of the conformal field theory evaluated at $q=e^{-2\pi t}$
can be thought of as the vacuum expectation values of a torus with
cycles $1$ and $it$. Under the $S$ transformation $t\mapsto 1/t$
the two cycles of the torus are interchanged. In order to describe the
resulting torus in terms of a torus with cycles $1$ and $i\tilde{t}$,
we then have to rescale the whole torus by $t^{-1}$. The   field
theory we are considering here is not conformally invariant since it
contains a mass term. However, it is conformally covariant in the
sense that it is invariant under rescalings provided that the mass
term is scaled appropriately. The above rescaling then requires that
we replace $m$ by $\widehat{m}=mt$.

\subsection{Cylinder diagrams of identical D-brane instantons}

We now have the machinery needed to analyse the
various modular conditions  that relate closed and open strings.
Here we will begin by considering the cylinder diagram joining
two identical D-branes, or a D-brane and its anti-brane.  In this
subsection all our considerations apply to the case of
D$(p+1)$-instantons.  The corresponding analysis for D$p$-branes
is a straightforward extension of this and will be described later.

The closed-string boundary states contain normalisation factors
that are easiest to establish by first evaluating the diagram in
the open-string channel in which case the diagram is a trace
over open-string states. The analogue of (\ref{openstring}) is now
\be
\tilde{Z}_{Dp;\bar Dp} (\tilde{t}) =
(2\sinh \pi m)^{3-p} \;
\left({f_4^{(\widehat{\m})}(\tilde{q}) \over
f_1^{(\widehat{\m})}(\tilde{q})}\right)^8\,,
\label{onelooppp}
\ee
for the brane-antibrane case, and
\be
\tilde{Z}_{Dp;Dp} (\tilde{t}) =
 (2\sinh \pi m)^{3-p}  \;
\left({f_1^{(\widehat{\m})}(\tilde{q}) \over
f_1^{(\widehat{\m})}(\tilde{q})}\right)^8
= (2\sinh \pi m)^{3-p} \,,
\label{loopbps}
\ee
for the brane-brane case. Here $\tilde q=e^{-{2\pi/t}}$, we have used
(\ref{mhat}), and $\bar Dp$ denotes the anti-brane corresponding to
$Dp$. There is no dependence on the transverse coordinates since the
supersymmetry conditions require that both branes are at $\x=0$.
In order to move the branes to non-zero transverse positions it is
necessary to turn on the flux of the Born--Infeld world-volume field.
Since the Born--Infeld flux is zero in the ansatz that we made for the
boundary states it is unsurprising that the cylinder amplitude does
not have sensible behaviour when the branes are placed away from
$\x=0$. The numerator factors
$(f_4^{(\widehat{\m})}(\tilde{q}))^8$ and
$(f_1^{(\widehat{\m})}(\tilde{q}))^8$ come from the fermionic part
of the trace while the denominator
factor $(f_1^{(\widehat{\m})}(\tilde{q}))^{-8}$ comes from the bosonic
part. In either case the trace over the non-zero modes is quite
standard but the treatment of the zero modes differs from the case of
flat space.  In particular, there is no power of
$\tilde{t}^{-{(p+1)/2}}$ from momentum integration along the
Neumann directions. This is in accord with the fact that in the
$pp$-wave background, the transverse momentum is not
continuous.
Instead, the contribution of the bosonic zero modes is
given by the factor
\be\label{openzero}
\tilde{Z}^0_{Dp;\bar Dp} = {\tilde{q}^{\,{p+1\over 2}\widehat{m}} 
\over (1-\tilde{q}^{\,\widehat{m}})^{p+1}} \,,
\ee
due to the contribution of $(p+1)$ harmonic oscillators \cite{dab}.
In the brane-brane case, there are eight fermionic zero modes
that combine to give four creation operators that raise the energy by
$\widehat{m}$, as well as four annihilation operators that lower the
energy by $\widehat{m}$. The trace over these creation operators then 
gives rise to the factor 
$\tilde{q}^{\,-2 \widehat{m}}\, (1-\tilde{q}^{\,\widehat{m}})^4$. 
Together with (\ref{openzero}) the total zero mode contribution is
therefore $(2\sinh \pi m)^{3-p}$ as in (\ref{loopbps}).  The
limit $\widehat{m}\rightarrow 0$ is in general singular, and does not
simply reproduce the flat space result.

For the case of a brane and an anti-brane, there are no fermionic zero
modes, as is reflected in the function
$(f_4^{(\widehat{\m})}(\tilde{q}))^8$. The bosonic zero mode
contribution $(1-\tilde{q}^{\,\widehat{m}})^{-p-1}$ in (\ref{openzero})
is then reproduced by  
$(2\sinh \pi m)^{3-p}$ as well as the factor of
$(1-\tilde{q}^{\,\widehat{m}})^{-4}$ that is contained in 
$(f_1^{(\widehat{\m})}(\tilde{q}))^{-8}$.
The ground state energy of the open string in (\ref{onelooppp}) is 
\be\label{groundenergy}
{p-3 \over 2}\, \widehat{m}  + 8 \Delta_{\widehat{m}} 
- 8 \Delta'_{\widehat{m}}\,.
\ee
The Casimir energy $8\Delta_{\widehat{m}}$ contains a contribution
from the bosonic zero modes that is equal to $2\widehat{m}$. Together
with the first term in (\ref{groundenergy}), this reproduces the
ground state energy of (\ref{openzero}). (The other terms in
(\ref{groundenergy}) describe the contribution to the ground state
energy coming from the non-zero modes.)

Now we turn to checking the consistency condition (\ref{cftcond}).
Using (\ref{beautiful}), the expressions (\ref{onelooppp}) and
(\ref{loopbps}) translate into the closed-string overlaps,
\ba\label{overlapone}
\tilde{\A}_{Dp;\bar Dp}(t) & = &  (2\sinh \pi m)^{3-p}
\left({f_2^{(\m)}(q) \over f_1^{(\m)}(q)}\right)^8\,, \\
\tilde{\A}_{Dp;Dp}(t) & =&   (2\sinh \pi m)^{3-p} \,
\left({f_1^{(\m)}(q) \over f_1^{(\m)}(q)}\right)^8 =
 (2\sinh \pi m)^{3-p} \,, \nn
\ea
where $q=e^{-2\pi t}$. The consistency condition is now that these
expressions are reproduced by a boundary state calculation in
the closed-string channel as in (\ref{Atilde}). The factor
$(2 \sinh \pi m)^{3-p}$ 
can be absorbed into the
normalisation of the boundary states. The contribution of the non-zero
fermionic and bosonic modes is again easily seen to reproduce the
relevant contributions to the numerator and denominator factors. The
only issue is whether the contribution of the `zero modes' to the
amplitude produces the correct $q$-dependent prefactors. Let us first
analyse the bosonic zero mode contribution. For each of the eight
directions, the contribution to the amplitude is given by
\ba
{}_b\langle 0 | e^{\pm{1 \over 2} \bar{a}_0 \bar{a}_0} \,
q^{\half \m a_0 \bar{a}_0}\, e^{\pm{1\over 2} a_0 a_0 } | 0 \rangle_b
& =& \sum_{r=0}^{\infty} {1\over r!\, r!}\, {1\over 2^{2r}} \,
{}_b\langle 0 | (\bar{a}_0 \bar{a}_0)^r (a_0 a_0)^r |0\rangle_b \,
q^{r\m} \nn\\
& = &\sum_{r=0}^{\infty} {2r \choose r}\, {1\over 2^{2r}}\, q^{r\m} \nn\\
& =& (1-q^\m)^{-\half}\,,
\label{boszerocal}
\ea
where the sign $\pm$ in the exponents depends on whether the direction
is Dirichlet or `Neumann'. The eight directions together therefore
produce the prefactor $(1-q^\m)^{-4}$, as required in
(\ref{overlapone}).

The contribution of the fermionic zero modes depends sensitively
on the value of $p$.  Only specific values of $p$ result in the
correct contributions as deduced from the open-string channel, namely,
a factor of $(1-q^\m)^4$  if both D-branes are of the same type and of
$(1+q^\m)^4$ if one brane is the anti-brane of the other.  These
values of $p$ coincide with those of the supersymmetric D-branes of
\cite{bp}. To see this, we shall analyse the possible values of $p$ in
turn.

The simplest example is the (non-supersymmetric) euclidean D3-brane
(the D$4$-instanton) for which the Neumann directions are chosen to be
$x^1,x^2,x^3,x^4$. Given the discussion of section~\ref{boundstat},
the fermionic ground state of this boundary state is annihilated (for
$\eta=+1$) by $\bar\theta^a_L$ and $\theta^a_R$.  The ground state has
eigenvalue zero under the
action of the fermionic zero mode part of the hamiltonian
(\ref{lcham})  and
therefore the contribution to (\ref{overlapone}) is zero for the case
where there are two branes, and $16$ if one brane is the anti-brane of
the other. In neither case does it give the correct factor of
$(1\pm q^\m)^4$ and thus the boundary state is inconsistent.

The analysis is similar for the case of the D$(-1)$-brane (the
D-instanton) for which the fermionic ground state of the boundary
state is annihilated (for $\eta=+1$) by $\theta^a_L$ and
$\theta^a_R$. Now the ground state has eigenvalue $2m$ under the
action of the fermionic zero mode part of the hamiltonian, and
therefore the contribution to (\ref{overlapone}) is zero for the case
of two branes, and $16 q^{2m}$ if one of the two branes is an
anti-brane. Therefore, the corresponding boundary state is again
inconsistent.

The situation is different for the supersymmetric  euclidean
D1-brane (the D2-instanton) whose
world-volume extends along the $x^1,x^2$ directions. It is not
difficult to check that its ground state is annihilated (for
$\eta=+1$) by
\be
\left( \bar\theta^a_L + (\gamma^1 \gamma^2)_{ab}\, \theta^b_L \right)
|\!| D1 \rangle\!\rangle^0 = 0\, , \qquad
\left( \theta^a_R - (\gamma^1 \gamma^2)_{ab}\, \bar\theta^b_R \right)
|\!| D1 \rangle\!\rangle^0 = 0\,.
\label{D1zero}
\ee
Thus, we can write the fermionic ground state component of the
D1-brane boundary state as
\be\label{D1ground}
|\!| D1 \rangle\!\rangle^0 = \exp \left(
\half (\gamma^1\gamma^2)_{ab} \bar\theta^a_R \bar\theta^b_R
- \half (\gamma^1\gamma^2)_{ab} \theta^a_L \theta^b_L \right)
|0\rangle_f \,,
\ee
where $|0\rangle_f$ denotes the lowest energy state that is
annihilated by $\bar\theta^a_L$ and $\theta^a_R$. The contribution of
the fermionic zero modes to the cylinder diagram can now be evaluated
in a straightforward manner.  Recalling from (\ref{lcham})
that $iHX^+$ contains the zero mode fermions in the term
$S_0^a\,\Pi_{ab}\, \tilde S_0^b$
we may expand the overlap $\langle Dp'|
e^{iHX^+}|Dp\rangle$  in an infinite power series in
$S_0^a\, \Pi_{ab}\, \tilde S_0^b$.  Details are given in
appendix~\ref{pedes} where the result is shown to be the expected
factor of $(1-q^\m)^4$ in the case of two branes, and $(1+q^\m)^4$
in the case when one brane is the anti-brane of the other. This
analysis is essentially identical for the case of pairs of
euclidean D5-branes (D6-instantons) of \cite{bp} which fill up the
directions $x^1, \dots, x^6$.

For the euclidean D3-brane that stretches, for example,
along $x^1,x^2,x^3,x^5$, the analogue of (\ref{D1zero}) is
\be
\left( \bar\theta^a_L - (\gamma^1 \gamma^2\gamma^3 \gamma^5)_{ab}\,
\bar\theta^b_R \right) |\!| D3 \rangle\!\rangle^0 = 0\, , \qquad
\left( \theta^a_R + (\gamma^1 \gamma^2\gamma^3\gamma^5)_{ab}\,
\theta^b_L \right) |\!| D3 \rangle\!\rangle^0 = 0\,,
\label{D3zero}
\ee
from which one can similarly show that the contribution of the
fermionic zero modes to the cylinder diagram is the required factor of
$(1\pm q^\m)^4$.

The other cases can be treated similarly. In summary, we have found
that the boundary states for which the cylinder diagram between
identical instantonic D-branes has consistent open-string and
closed-string interpretations correspond precisely to the
supersymmetric cases found in \cite{bp}.

\subsection{Cylinder diagrams for pairs of different D-brane instantons}

In the previous subsection we have seen that the supersymmetric
D$(p+1)$-instantons that were described in \cite{bp}
satisfy the consistency condition (\ref{cftcond}) provided
the normalisation  of the D$(p+1)$-instanton
boundary state  is
\be\label{boundarynorm}
{\cal N}_{D(p+1)} = 
\left(2 \sinh \pi m \right)^{{3-p\over 2}}\,.
\ee
We can now check whether, with this normalisation, the cylinder
diagrams joining different instantonic D-branes also satisfy the
open-closed consistency condition. Let us discuss a few cases
explicitly.

For the supersymmetric
configuration of two orthogonal euclidean D1-branes, one stretching
along $x^1,x^2$, and one along $x^3,x^4$, we find
\be
\tilde{\A}_{D1;D1'} =(2\sinh \pi m)^2\;
\left( {f_1^{(m)}(q)   f_2^{(m)}(q) \over f_1^{(m)}(q)   f_2^{(m)}(q)}
\right)^4  = (2\sinh \pi m)^2 \,.
\label{D1D1p}
\ee
Under the $S$ transformation this gives rise to
\be
\tilde{Z}_{D1;D1'} =
(\tilde{q}^{\,-\widehat{m}/2}- \tilde{q}^{\,\widehat{m}/2})^{2} \;
\left({f_1^{(\widehat{\m})}(\tilde{q}) f_4^{(\widehat{\m})}(\tilde{q})
\over f_1^{(\widehat{\m})}(\tilde{q}) f_4^{(\widehat{\m})}(\tilde{q})
}\right)^4 = 
\tilde{q}^{\,-\widehat{m}} \, (1- \tilde{q}^{\,\widehat{m}})^{2}  \,,
\label{D1D1popen}
\ee
which reproduces the one-loop open-string calculation.  In particular,
in this case there are no $NN$ directions and
the contribution of the bosonic zero modes arising from the
numerator factor of $f_1^{(\widehat{\m})}(\tilde{q})^{-4}$ cancels
the factor of $(1-\tilde{q}^{\,\widehat{\m}})^{2}$.

For the configuration of two euclidean D$1$-branes, one stretching
along $x^1,x^2$, and one along $x^1,x^3$, the result is
\be
\tilde{\A}_{D1;D1''} =   (2\sinh \pi m)^2 \,
{ \left(f_2^{(m)}(q^2)\right)^4 \over
\left(f_1^{(\m)}(q)\right)^6\,  \left(f_2^{(\m)}(q)\right)^2}
\label{D1D1pp}
\ee
instead of (\ref{D1D1p}). Under the $S$ transformation this
becomes
\be\label{D1D1ppopen}
\tilde{Z}_{D1;D1''} =  \tilde{q}^{\,-\widehat{m}} \,
(1 - \tilde{q}^{\,\widehat{m}})^{2} 
{ \left(f_4^{(2\widehat{m})}(\tilde{q}^{\half})\right)^4 \over
\left(f_1^{(\widehat{m})}(\tilde{q})\right)^6\,
\left(f_4^{(\widehat{m})}(\tilde{q})\right)^2}\,.
\ee
The net contribution of the bosonic zero modes in the open-string
picture is then proportional to $(1-\tilde{q}^{\,\widehat{\m}})^{-1}$,
as is appropriate for a single $NN$ direction. The numerator describes
the contribution coming from the open-string fermions with mass
$\widehat{m}$, since
\be
f_4^{(2\widehat{m})}(\tilde{q}^{\half}) =
\tilde{q}^{-\half \Delta^\prime_{2\widehat{m}}}\,
\prod_{n=1}^{\infty}
\left(1 - \tilde{q}^{\sqrt{\widehat{\m}^2+(n/2-1/4)^2}}\right)\,.
\ee\label{f4half}
The configuration of a euclidean D$1$-brane stretching along
$x^1,x^2$, and a D$3$-brane stretching along $x^1,x^2,x^3,x^5$ gives
\be
\tilde{\A}_{D1;D3} =  \tilde{q}^{\,-\widehat{m}/2}\,
(1 - \tilde{q}^{\,\widehat{m}})\,
{ \left(f_2^{(m)}(q^2)\right)^4 \over
\left(f_1^{(\m)}(q)\right)^6\,  \left(f_2^{(\m)}(q)\right)^2}\,,
\label{D1D3}
\ee
since the normalisation of the D3-brane boundary state does not
involve any power of 
$2\sinh \pi m$. 
The contribution of the bosonic 
zero modes in the open-string picture is then proportional to
$(1-\tilde{q}^{\,\widehat{\m}})^{-2}$, as is appropriate for a
configuration with two $NN$ directions.

Finally, the configuration of a D1-brane stretching along $x^1,x^2$,
and a D3-brane stretching along $x^1,x^3,x^4,x^5$ gives
\be
\tilde{\A}_{D1;D3'} =   (2\sinh \pi m)\,
\left({f_1^{(\m)}(q) f_2^{(\m)}(q) \over
f_1^{(\m)}(q) f_2^{(\m)}(q)}\right)^4 =
\tilde{q}^{\,-\widehat{m}/2}\,
(1 - \tilde{q}^{\,\widehat{m}})\,.
\label{D1D3p}
\ee
In this case the contribution of the bosonic zero modes in the open
string picture is proportional to $(1-\tilde{q}^{\,\widehat{\m}})^{-1}$,
as is appropriate for a configuration with a single NN direction.
All other configurations can be treated in a similar fashion.

\subsection{Some comments about supersymmetry constraints}

As seen in the previous two subsections, the conformal field theory
condition (\ref{cftcond})  singles out those boundary states
that correspond to the D$(p+1)$-instantons that were found in
\cite{bp}. All other states violate this condition and the
corresponding instantonic D-branes are therefore probably
inconsistent.

The condition  (\ref{cftcond}) is unambiguously defined for a
cylindrical  world-sheet of fixed  ratio of circumference to length,
$2\pi p^+/X^+$. In applying this formula to  D-branes separated at
fixed values of $x^-$ we need to Fourier transform the overlap with
respect to $p^+$ to obtain a position-space function defined by the
integral (\ref{covariant}).  In fact, this integral is
divergent for any pair of D$(p+1)$-instantons in the $pp$-wave
background. We
have not understood the  physical interpretation of such
divergences.

An unusual feature of the expressions we have obtained is that the
interaction `energy' (more properly, action) between identical
supersymmetric D$(p+1)$-instantons does not vanish (see
(\ref{overlapone}) above).
Presumably this means that the combined system
of supersymmetric D-branes is not   BPS in the usual sense.
Instead, for constant $p^+ > 0$ the overlaps
are independent of $X^+$.  These
properties are consistent with the supersymmetry algebra
in the type IIB $pp$-wave background.
Recall that in the flat case, one can use the fact that the
BPS boundary states satisfy (\ref{kinematical}) together with the
anti-commutation relations of the kinematical supercharges,
$\{Q_a,Q_b\}\sim p^+$, to deduce that the BPS-BPS amplitude $\tilde\A$
must vanish. This follows simply from the relation (written
symbolically)
\be
{\cal A} =  \langle\!\langle Dp_1  |\!|\;  e^{-H X^+}\; |\!|  Dp_2
 \rangle\!\rangle =
{1\over p^+}\; \langle\!\langle Dp_1  |\!|\;  e^{-H X^+}\;
Q^a \tilde Q^a\; |\!| Dp_2 \rangle\!\rangle =0 \,,
\label{susykin}
\ee
where the properties (\ref{kinematical}) of the boundary states have
been used.  This argument
relies on the fact that the kinematical supercharges commute with $H$.
However, in the $pp$-wave background
the kinematical supercharges, $Q^a$ and $\tilde Q^a$, do not
commute with $H$ so the overlap cannot be argued to vanish in this
manner. The algebra of the dynamical supercharges is slightly less
constraining in the case of a flat background since it implies
that
\be
\langle\!\langle Dp_1  |\!|\;  e^{-H X^+}\;
Q^{\dot a} \tilde Q^{\dot a}\; |\!|
Dp_2 \rangle\!\rangle
= \langle\!\langle Dp_1  |\!|\;  e^{-H X^+}\;H\;
|\!|  Dp_2 \rangle\!\rangle = 0\, ,
\label{susydyn}
\ee
which means that ${\partial \over \partial X^+} \, {\cal A} =0$.
This argument also applies to the $pp$-wave case since the
dynamical supercharges do commute with $H$
and the $\mu$-dependent correction terms in the anti-commutator do not
contribute to the overlap.
Consequently, the overlaps between pairs of identical D-branes
must be independent of $X^+$,
as is indeed consistent with (\ref{overlapone}).  This explains
why ${\cal A}$ is determined equally well  in the long cylinder
($X^+ \to \infty$) or short cylinder ($X^+ \to 0$) limit.
From the open-string point of view, the non-vanishing of the
brane-brane  amplitude follows from the fact that there is a single
(bosonic) ground state. Furthermore, since the excited open-string
levels have equal numbers of bosonic and fermionic states,  the
amplitude is independent of $q$, and therefore independent of $X^+$.
This is an example of a non-zero contribution to the Witten index.

\section{Time-like branes}

For lorentzian signature world-volumes
the $x^\pm$ coordinates are directions in the world-volume of the
brane and the light-cone gauge can be chosen in the open-string sector
in the usual manner.  The following discussion will be somewhat formal
since  the boundary states that
we have used restrict all D-branes to the same transverse position
$x^I$ =0, where $I= p, \dots, 8$.   Therefore the integrated
 expression for the cylinder amplitude joining two D-branes will
be  singular since the branes are coincident.
The D-branes could, in principle, be separated by turning
on a flux of the Born--Infeld field but we have not considered
this situation here.  We may still hope to determine a consistency
condition such as (\ref{cftcond}) since this  relates the
integrands in the closed-string and open-string channels rather than
the divergent integrals.

In order to motivate the argument we shall again review the
expressions for the cylinder amplitudes for separated D-branes
in flat space, and generalize to the $pp$-wave background at the end.
For large enough transverse separation the amplitude is given by the
integral of a trace  over the open-string states,
\ba
Z_{Dp;Dp}(\x_1;\x_2) &=&
\int_0^\infty {dt\over t} \int {dp^+ dp^-\over 2\pi} \; \Tr\;
e^{- ( p^+ t H^{open} + t p^+ p^-)}
\nn\\
& = & \int_0^\infty {dt\over t} \int {dp^+ dp^-\over 2\pi} \;
e^{- t p^+ p^-}
\;
\tilde{Z}_{Dp;Dp}(t,{\bf x}_1,{\bf x}_2) \,,
\label{ttime}
\ea
where $t = X^+/2\pi p^+$, and
\be\label{Ztildedef}
\tilde{Z}_{Dp;Dp}(t,{\bf x}_1,{\bf x}_2) = \Tr\;
e^{-   p^+  t H^{open}}\,.
\ee
The trace includes integration over the momenta $k^I$
($I=1,\ldots,p-1$) in the Neumann directions.
In the flat space case
\be\label{lchamopen}
p^+ H^{open} =
{1\over 2\pi} \sum_{I=p}^8 (x_1^I-x_2^I)^2 + \pi
\sum_{I=1}^{p-1} (k^I)^2 + 2 \pi N^{open}\,,
\ee
where $N^{open}$ is again the total level number of the bosonic and
fermionic open-string oscillators, including the usual normal ordering
constants. In the $pp$-wave case, $H^{open}$ is
replaced by the formula described in \cite{dab}. In particular, the
position and momentum-dependence is then described by the zero mode
term $m \sum_{I=1}^{p-1} a_0^I \bar a_0^I$.  A Wick rotation,
$p^0 \to i p^0$ replaces $p^+$ and $p^-$ in (\ref{ttime})
by the complex variables $p=(p^9 + i p^0)/\sqrt 2$ and
$\bar p = (p^9 - ip^0)/\sqrt 2$ so the exponent in the
integrand becomes $t p \bar p$.
Performing the $p$ and $\bar p$ integrations gives
\be
Z_{Dp;Dp} (\x_1;\x_2)=
\int_0^\infty {dt\over t^2} \;
\tilde{Z}_{Dp;Dp}(t,{\bf x}_1,{\bf x}_2)\,.
\label{openone}
\ee
In order to describe this process from the closed-string point of view
we need to consider a non-standard light-cone gauge in the closed
sector in which the r\^oles of $X^+$ and $p^+$ are reversed.  In this
case the cylinder interaction is given in terms of the overlap
\be\label{realcyl}
\A_{Dp;Dp}(\x_1,\x_2) =
\langle\!\langle Dp, \x_1 |\!|
          {1\over   X^+ H} |\!| Dp, \x_2 \rangle\!\rangle \,.
\ee
Here the closed-string propagator has been written as
$p^+ p^- + X^+ H$, where $X^+ H$ is given by the same expression
as $p^+ H$ in (\ref{lcham}), or by the usual hamiltonian in the flat
space case. In deriving (\ref{realcyl}) we have used that the
boundary states are (formally) annihilated by $p^+$ and $p^-$ since
they satisfy Neumann boundary conditions along the light-cone
directions. Writing $1/X^+ H$ in terms of an integral gives
\be\label{b1}
\A_{Dp;Dp}(\x_1,\x_2) =
\int_0^\infty d\tilde{t} \,
   \langle\!\langle Dp, \x_1 |\!|
                e^{-X^+ H \tilde{t}} |\!| Dp,\x_2\rangle\!\rangle
   = \int_0^\infty d\tilde{t}\,
\tilde{\A}_{Dp;Dp}(\tilde{t},\x_1,\x_2) \,.
\ee
As before the integration variable $\tilde{t}$ is related to $t$ in
(\ref{openone}) above as $\tilde{t}=1/t$. Using this substitution we
thus obtain
\be\label{b2}
\A_{Dp;Dp}(\x_1,\x_2) =  \int_0^\infty
{dt \over t^2} \;
\tilde{\A}_{Dp;Dp}(t^{-1},\x_1,\x_2) \,.
\ee
The two expressions (\ref{b2}) and (\ref{openone}) therefore agree
provided that $\tilde{\A}$ and $\tilde Z$ satisfy (\ref{cftcond}) with
$\tilde{Z}(t,\x_1,\x_2)$ being given by
(\ref{Ztildedef}). Thus the consistency condition reduces again to
the conformal field theory condition (\ref{cftcond}). The analysis of
the consistent time-like D$p$-branes is then identical to that
for the case of the D$(p+1)$-instantons before. In particular, the
only consistent values of $p$ correspond to those of the
supersymmetric D$p$-branes that were found in \cite{dab}.

\section{Conclusion}

In this paper we have analysed the open-closed consistency condition
for the D$(p+1)$-instantons and D$p$-branes in the $pp$-wave background.
The D-branes that are consistent are those
that are known to preserve half of the supersymmetries \cite{bp,dab}.
Our analysis relies on the striking identities
(\ref{beautiful}) that describes the $S$ modular transformation
relations for the characters in the $pp$-wave conformal field theory.

It is straightforward to identify the quarter-BPS D-branes proposed in
\cite{ST} based on a supergravity analysis.  These are states that
preserve only eight components of supersymmetry, defined by the
conditions  on the Killing spinors given by equations (6.35) and
(6.36) of 
\cite{ST}.\footnote{The only exception seems to be the D1-brane for
which the world-volume lies along the light-cone directions.}
These are precisely the components that correspond to the linear
combination of `kinematical' supersymmetries that entered in
(\ref{kinematical}).   In other
words, the boundary states corresponding to these D-branes are not
required to satisfy (\ref{dynamical}).  However, our analysis
suggests that such D-branes do not generally satisfy the
consistency condition that relates the closed-string and
open-string sectors.  The condition (\ref{dynamical}) seems to be
crucial for ensuring such consistency.

The null branes that are identified with giant gravitons at fixed $x^-$
\cite{maldetal,ST} are not seen with the choice of light-cone
gauge that we have used although they should obviously exist as
consistent D-branes.

We have seen that the interactions between identical supersymmetric
D-branes do not vanish and that this is in accord with expectations
based on the supersymmetry algebra.  This seems to be connected with
the observation that the $pp$-wave background exhibits `tidal forces'
which will cause D-branes to repel each other\footnote{We are grateful
to Gary Gibbons for discussions on this point.}.

\section*{Acknowledgements}

We thank Costas Bachas, Gary Gibbons and David Mateos for useful
conversations. 
This work was begun while the authors were visiting the Isaac Newton
Institute in Cambridge. OB is supported in part by the Israel Science
Foundation under grant no.~101/01-1. MRG is grateful to the Royal
Society for a University Research Fellowship. We also acknowledge
partial support from the PPARC Special Programme Grant `String Theory
and Realistic Field Theory', PPA/G/S/1998/0061 and the EEC contract
HPRN-2000-00122.

\appendix

\section{Derivation of the $S$ modular transformation formula}
\label{deriv}

In this appendix we will derive the modular transformation identities
(\ref{beautiful}). We shall only give the argument for the case of the
transformation of $f_1^{(\m)}(q)$ in detail since the arguments for
the case of $f_2^{(\m)}(q)$ and $f_3^{(\m)}(q)$ are very similar.
Somewhat related formulae have been derived, using different
techniques, in \cite{salitz}\footnote{We thank Andreas Recknagel for
drawing our attention to this paper.}. The expression for the
logarithm of (\ref{fdef}) is
\ba
\log f_1^{(\m)}(q) & =&  2 \pi t \Delta_\m + \half \log (1-q^\m)
+ \sum_{n=1}^{\infty} \log \left(1-q^{\sqrt{\m^2+n^2}}\right) \cr
& = & 2 \pi t \Delta_\m + \half \sum_{n\in\Zop}
\log \left(1-q^{\sqrt{\m^2+n^2}}\right) \cr
& = & 2 \pi t \Delta_\m - \half \sum_{n\in\Zop} \sum_{p=1}^{\infty}
{1\over p}\, q^{p\, \sqrt{\m^2+n^2}} \,.
\label{derone}
\ea
 Next we use the identity (see \cite{gr} (8.432:6)
and (8.469:3))
\be
e^{-z} = {1\over \sqrt{\pi}} \int_0^\infty dr\, r^{-1/2}
e^{-r-{z^2\over 4r}} \,,
\label{bessel}
\ee
to write $q^{p\, \sqrt{\m^2+n^2}}$ in the last line of (\ref{derone})
in terms of an integral. This gives
\be
\log f_1^{(\m)}(q) =
 2 \pi t \Delta_\m - {1\over 2 \sqrt{\pi}} \sum_{n\in\Zop}
\sum_{p=1}^{\infty}
\int_0^{\infty} ds\, s^{-1/2} e^{-p^2 s - \pi^2 t^2 (\m^2+n^2)/s} \,,
\label{dertwo}
\ee
after substituting $r=p^2 s$. Using the Poisson resummation
formula
\be
\sum_{n\in\Zop} e^{-\alpha\pi n^2} = \alpha^{-1/2}
\sum_{\hat{n}\in\Zop} e^{-\pi \hat{n}^2 / \alpha} \,,
\label{poisson}
\ee
the sum over $n$ in (\ref{dertwo}) can be rewritten as
\ba
\log f_1^{(\m)}(q)   &= &  2 \pi t \Delta_\m
- {1\over \pi t} \sum_{\hat{n}=1}^\infty \sum_{p=1}^{\infty}
\int_0^\infty ds\, e^{-p^2 s} e^{-\pi^2 \m^2 t^2 /s}
e^{-\hat{n}^2 s / t^2} \nn\\
&& - {1\over 2 \pi t} \sum_{p=1}^{\infty} \int_0^\infty ds \,
e^{-p^2 s} e^{-\pi^2 \m^2 t^2 /s} \,.
\label{derthree}
\ea
The last line of (\ref{derthree}) is equal to
$2\pi \Delta_{\widehat{\m}} / t$. The sum in the first line
can be extended
to a sum over all integers $p$ except for a correction term
corresponding to $p=0$,
\ba
\log f_1^{(\m)}(q)   &= &  2 \pi t \Delta_\m +
{2\pi \over t} \Delta_{\widehat{\m}}
- {1\over 2 \pi t} \sum_{\hat{n}=1}^\infty \sum_{p\in\Zop}
\int_0^\infty ds\, e^{-p^2 s} e^{-\pi^2 \m^2 t^2 /s}
e^{-\hat{n}^2 s / t^2} \nn\\
&& + {1\over 2 \pi t} \sum_{\hat{n}=1}^\infty
\int_0^\infty ds\, e^{-\hat{n}^2 s / t^2} \,
e^{-\pi^2 \m^2 t^2 /s} \,.
\label{derfour}
\ea
Upon substituting $\bar{s} = s/t^2$ in the last term in (\ref{derfour}) it
becomes precisely $-2\pi t \Delta_{\m}$, and thus cancels the first
term. Now the Poisson resummation formula can be used for the sum over
$p$ giving,
\ba
\log f_1^{(\m)}(q) & =&  {2\pi \over t} \Delta_{\widehat{\m}}
- {1\over 2 \sqrt{\pi} t} \sum_{\hat{n}=1}^\infty
\sum_{\hat{p}\in\Zop}
\int_0^\infty ds\, s^{-1/2} \,
e^{-\pi^2 \hat{p}^2/ s} e^{-\pi^2 \m^2 t^2 /s}
e^{-\hat{n}^2 s / t^2} \nn\\
& =&  {2\pi \over t} \Delta_{\widehat{\m}}
- {1\over 2 \sqrt{\pi}} \sum_{\hat{n}=1}^\infty {1\over \hat{n}}
\sum_{\hat{p}\in\Zop}
\int_0^\infty d\bar{s}\, \bar{s}^{-1/2} \, e^{-\bar{s}}
e^{-\pi^2 \hat{n}^2 (\hat{p}^2+\widehat{\m}^2)/ t^2 \bar{s}} \,,
\label{derfive}
\ea
where we have substituted $\bar{s} = \hat{n}^2 s / t^2$ in the last
line and used again $\widehat{\m}=\m\, t$. Finally we use
(\ref{bessel}) again to rewrite this as
\ba
\log f_1^{(\m)}(q) & =&  {2\pi \over t} \Delta_{\widehat{\m}}
- \half \sum_{\hat{p}\in\Zop}
\sum_{\hat{n}=1}^\infty {1\over \hat{n}}
e^{-2\pi \hat{n} \sqrt{\hat{p}^2+\widehat{\m}^2}/t} \nn\\
& =&  {2\pi \over t} \Delta_{\widehat{\m}}
+ \half \sum_{\hat{p}\in\Zop}
\log \left( 1 - \tilde{q}^{\sqrt{\hat{p}^2+\widehat{\m}^2}} \right)
\nn\\
& =& \log f_1^{(\widehat\m)}(\tilde{q}) \,,
\label{dersix}
\ea
where $\tilde{q}=e^{-2\pi/t}$.

\section{Fermionic ground state contribution to cylinder diagram}
\label{pedes}

In this appendix we shall give a pedestrian derivation of the
contribution of the fermionic zero modes to the cylinder diagram of
two identical euclidean D1-branes. All other cases can be obtained by
obvious generalizations.
We will choose the orientation of the world-sheet so that
$\epsilon_1=\epsilon_2=-1$, with all other $\epsilon_I$ equal to
$+1$. Given the boundary conditions satisfied by $S_0^a$ and $\tilde S_0^a$
(\ref{boundarycondii}) the expansion of the action of the zero-mode fermion
bilinears satisfies the following equation
\be
\Bigl(i\, \m\, S^a_0\, \Pi_{ab} \tilde{S}^b_0\Bigr)^n \,
|\!| D1\, \rangle\!\rangle_0 =  (-1)^n \eta^n \, \m^n \,
\prod_{l=1}^{n} (\gamma^3 \gamma^4)_{a_l b_l}
\tilde{S}^{b_1}_0 \cdots \tilde{S}^{b_n}_0 \tilde{S}^{a_n}_0\cdots
\tilde{S}^{a_1}_0 |\!| D1\, \rangle\!\rangle_0 \,.
\label{fermcalctwo}
\ee
In the tree level diagram with the D1-brane boundary state, only the
even powers of $n$ contribute. After computing the appropriate
$\gamma$-matrix traces we find that on the bivector states the $n=2l$
contribution (with $l\geq 1$) is
\be
\Bigl(i\, \m\, S^a_0\, \Pi_{ab} \tilde{S}^b_0\Bigr)^{2l} \, M_{IJ}
|I\rangle |J\rangle =
\half (4\m)^{2l} \epsilon^{34}_J M_{IJ} |I\rangle |J\rangle \,,
\label{fermcalcthree}
\ee
where $\epsilon^{34}_J=-1$ for $J=3,4$, and $+1$ otherwise. On the
bispinor states one finds on the other hand
\be
\Bigl(i\, \m\, S^a_0\, \Pi_{ab} \tilde{S}^b_0\Bigr)^{2l} \,
\left(\gamma^1\gamma^2\right)_{\dot{a} \dot{b}}
|\dot{a}\rangle |\dot{b}\rangle =
(2\m)^{2l}  \left(\gamma^1\gamma^2\right)_{\dot{a} \dot{b}}
|\dot{a}\rangle |\dot{b}\rangle \,.
\label{fermcalcfour}
\ee
Thus the fermionic zero mode contribution to the cylinder is
\ba
2 \, q^{2\m} \, \sum_{l=1}^{\infty} {1\over 2l!} (4 \pi t \m)^{2l}
& - &
8 \, q^{2\m} \, \eta \sum_{l=1}^{\infty} {1\over 2l!} (2 \pi t \m)^{2l}
+ 8 (1-\eta) q^{2\m}  \label{fermcalz}\\
& =& 16 \, q^{2\m} \left( {1\over 8} \cosh(4\pi t \m)
- \eta \, {1\over 2} \cosh(2\pi t \m) + {3\over 8} \right) \,,
\label{fermcalcfive}
\ea
where $\eta=+1$ for the overlap between two D1-branes, while
$\eta=-1$ for the overlap between a D1-brane and its anti-brane.
The prefactor $q^{2\m}$ comes from the constant term in
(\ref{lcham}), and the last term in (\ref{fermcalz}) comes from the
$l=0$ contribution.

For $\eta=+1$ (\ie\ for the brane-brane configuration) we now get,
using the standard product formula for the hyperbolic cosine (see for
example \cite{gr}),
\ba
 16 \, q^{2\m} \left( {1\over 8} \cosh(4\pi t \m)
- {1\over 2} \cosh(2\pi t \m) + {3\over 8} \right) & = &
16 q^{2\m}\, \sinh^4(\pi t \m) \cr
& = & q^{2\m} \left( q^{{\m\over 2}} - q^{-{\m\over 2}}\right)^4  \cr
& = & (1 - q^{\m})^4 \,,
\label{fermcalcsix}
\ea
as required. On the other hand, for $\eta=-1$, the $\sinh^4$ term
above is replaced by a $\cosh^4$ term, giving
$(1+q^{\m})^4$. This demonstrates that the D1-brane of
\cite{bp} gives rise to the correct overlap of boundary states,
(\ref{overlapone}), consistent with the open-string description.


\begin{thebibliography}{99}

\bibitem{hulletal}{M. Blau, J. Figueroa-O'Farrill, C. Hull,
G. Papadopoulos, {\it Penrose limits and maximal supersymmetry},
{\tt hep-th/0201081}.}

\bibitem{maldetal}{D. Berenstein, J. Maldacena, H. Nastase,
{\it Strings in flat space and pp waves from ${\cal N}=4$ Super Yang
Mills}, {\tt hep-th/0202021}.}

\bibitem{met}{R.R. Metsaev, {\it Type IIB Green-Schwarz superstring in
plane wave Ramond-Ramond background}, {\tt hep-th/0112044}.}

\bibitem{mt}{R.R. Metsaev, A.A. Tseytlin, {\it Exactly solvable model of
superstring in plane wave Ramond-Ramond background},
{\tt hep-th/0202109}.}

\bibitem{berk}{N. Berkovits, {\it Conformal field theory for the
superstring in a Ramond-Ramond plane wave background},
{\tt hep-th/0203248}.}

\bibitem{dab}{A. Dabholkar, S. Parvizi, {\it  Dp branes in PP-wave
background}, {\tt hep-th/0203231}.}

\bibitem{gg}{M.B. Green, M. Gutperle, {\it Light-cone supersymmetry and
D-branes}, Nucl. Phys. {\bf B476}, 484 (1996), {\tt hep-th/9604091}.}

\bibitem{bp}{M. Billo, I. Pesando, {\it Boundary states for GS
superstrings in an $Hpp$ wave background}, {\tt hep-th/0203028 v1}.}

\bibitem{gs1}{M.B. Green, J.H. Schwarz, {\it Supersymmetrical string
theories}, Phys. Lett. {\bf B109}, 444 (1982).}

\bibitem{ST}{K. Skenderis, M. Taylor, {\it Branes in AdS and pp-wave
spacetimes}, {\tt hep-th/0204054}.}

\bibitem{gone}{M.B. Green, {\it Pointlike structure and
off-shell dual strings},  Nucl. Phys. {\bf B124} 461 (1977);
M.B. Green, {\it The influence of world-sheet
boundaries on critical closed string theory}, Phys. Lett.
{\bf B302}, 29 (1993), {\tt hep-th/9212085}.}

\bibitem{gtwo}{M.B. Green, {\it Point-like states for type IIB
superstrings}, Phys. Lett. {\bf B329}, 435 (1994),
{\tt hep-th/9403040}.}

\bibitem{polcai}{J. Polchinski, Y. Cai, {\it Consistency of open
superstring theories}, Nucl. Phys. {\bf B296}, 91 (1988).}

\bibitem{gr}{I.S.Gradshteyn, I.M. Ryzhik, {\it Table of integrals,
series, and products}, Academic Press, 6th ed (2000).}

\bibitem{salitz} H. Saleur, C. Itzykson, {\it Two-dimensional field
theories close to criticality}. Journ. Stat. Phys. {\bf 48}, 449  
(1987).

\end{thebibliography}
\end{document}